\documentclass[%
 aip,
 amsmath,amssymb,
 reprint,%
floatfix]{revtex4-1}

\usepackage{graphicx}
\usepackage{dcolumn}
\usepackage{bm}

\usepackage[utf8]{inputenc}
\usepackage[T1]{fontenc}
\usepackage{mathptmx,color}

\begin{document}

\preprint{AIP/123-QED}

\title[On standardised moments of force distribution in simple liquids]{On standardised moments of force distribution in simple liquids}

\author{Jonathan Utterson}
\email{utterson@maths.ox.ac.uk}
\affiliation{Mathematical Institute, University of Oxford, Radcliffe Observatory Quarter, Woodstock Road, Oxford, OX2 6GG, United Kingdom}
\author{Radek Erban}%
\email{erban@maths.ox.ac.uk}
\affiliation{Mathematical Institute, University of Oxford, Radcliffe Observatory Quarter, Woodstock Road, Oxford, OX2 6GG, United Kingdom}

\date{\today}

\begin{abstract}
The force distribution of a tagged atom in a Lennard-Jones fluid in the canonical ensemble is studied with a focus on its dependence on inherent physical parameters: number density ($n$) and temperature ($T$). Utilising structural information from molecular dynamics simulations of the Lennard-Jones fluid, explicit analytical expressions for the dependence of standardised force moments on $n$ and $T$ are derived. Leading order behaviour of standardised moments of the force distribution are obtained in the limiting cases of small density ($n \rightarrow 0$) and low temperature ($T \rightarrow 0$), while the variations in the standardised moments are probed for general $n$ and $T$ using molecular dynamics simulations. Clustering effects are seen in molecular dynamics simulations and their effect on these standardised moments is discussed. 
\end{abstract}

\maketitle

\section{\label{sec:level1}Introduction}

\noindent
Understanding the moments and measures of a distribution for a fully atomistic molecular dynamics (MD) simulation allow us to better fit coarser models that reproduce these~\cite{Joshi:2020:RAC,Wang:2009,Erban:2020:SMR}. It is often the case in model coarse graining that we wish to directly reconcile the energy landscape of the fully atomistic system to a more basic representation that allows us to maintain as many physical properties of the system of interest, with as little computational cost as possible\cite{Ingolfsson:2014}. Though, it is also natural to match forces between the high and low resolution systems in an effort to reproduce the force distribution which will inherently give rise to the energy landscape~\cite{Davtyan:2015:DFM,Erban:2016:CAM,Wales:2018,Gunaratne:2019:SLI,Rolls:2017:VRR,Erban:2014:MDB}.

Let $\mathbf{F}=[F_1,F_2,F_3]$ denote a force on a tagged atom in a liquid. Depending on the relative positions of other atoms, force $\mathbf{F}$ can vary over a range of values and a detailed information on $\mathbf{F}$ can be obtained by calculating properties of its equilibrium distribution, which we will call force distribution in this manuscript. Considering an isotropic system, the equilibrium distribution of each force coordinate is the same. We define the standardised moment of the force distribution by averaging over the $k$-th power of its first coordinate as
\vskip -3.5mm
\begin{equation}
\label{statkurt}
\alpha_k
=
\frac{\left\langle 
F_1^k \right\rangle}{\left\langle F_1^2\right\rangle^{k/2}} \, ,
\end{equation}
\vskip -1mm
\noindent where $\left\langle 
F_1^k \right\rangle$ is the $k$-th moment of the force distribution and $\alpha_k$ standardises the $k$-th moment by scaling it with the $k$-th power of the standard deviation of the force distribution.
In a simple homogeneous fluid with radially symmetric interactions between particles, the force distribution will exhibit symmetry around the origin and thus all odd standardised moments vanish,
i.e.  $0 = \alpha_1 = \alpha_3 = \alpha_5 = \dots.$ 
As $\alpha_2 \equiv 1$ by definition~(\ref{statkurt}), the first non-trivial standardised moment is kurtosis, denoted $\alpha_4$, which provides a measure of spread that details how tailed the force distribution is relative to a normal distribution~\cite{DeCarlo:1997}. In this paper, we study how the force distribution depends on the number density of a homogeneous many-body system, and the temperature of the same system in a canonical ensemble. We will do this by studying the behaviour of the second moment of the force distribution $\left\langle F_1^2\right\rangle$
and standardised even moments $\alpha_4$, $\alpha_6,$ $\alpha_8$, $\dots$. 
If the force distribution was Gaussian, then the even standardised moments would be
\begin{equation}
\alpha_{k}
=
(k-1)!!
=
\prod_{i=1}^{k/2}
\, (2i-1),
\quad
\mbox{for}
\;
k=2,4,6,8,10,\dots, 
\;
\label{gaussmoments}    
\end{equation}
and the second moment $\left\langle F_1^2\right\rangle$ would be sufficient to parametrize the force distribution. However, the force distributions in simple liquids have been reported to deviate from Gaussian distribution~\cite{Shin:2010:BMM,Erban:2020:CMD,Carof:2014}. In particular, by comparing the results of our analysis with Gaussian moments in equation (\ref{gaussmoments}), we can also quantify how non-Gaussian the real force distribution is.

Much work has been done in the area of force distributions of many-body systems: with seminal work from Chandrasekhar~\cite{Chandrasekhar:1943} that employed Markov's theory of random flights to give an expression for the force distribution of a many-body system interacting through a $1/r$ gravitational potential. More recent work has been done with the help of MD by Gabrielli et al~\cite{Gabrielli:2006:FDR}, who derived an expression for the kurtosis of the force distribution  for a lattice system of atoms interacting through the gravitational potential. Further, using the classical density functional theory, an expression for the probability distribution of force for a system interacting through an arbitrary weakly repulsive potential was derived by Rickayzen et al~\cite{Rickayzen:2012:SPF,Branka:2011:PFD}. 
 
In this paper, we study the number density and temperature dependence of the force distribution for a many-body system interacting through a Lennard-Jones 12-6 potential~\cite{Jones:1924,Watanabe:2012}, which is ubiquitously used and has been shown to model homogeneous systems of interacting (Argon) atoms well~\cite{Rahman:1964:CMA,Verlet:1967,Hansen:1969}. 
 
In Section~\ref{1DTheory}, an in depth investigation is given to the simple two-body system in one spatial dimension, which provides the ideal platform to illustrate the underlying methods while retaining interesting dynamical behaviour. From first principles we derive first-order partial differential equations (PDEs) describing the dependence of the standardised moments of the force distribution has on parameters. In doing so we further derive an analytic expression for the partition function of a two-body system that depends solely on the standardised moments of the force distribution whereupon the expression is exact in an asymptotic limit of the density going to zero ($n \rightarrow 0$). Similarly, an expression is derived relating the average energy of the system to standardised moments of force from the temperature dependent PDE. In parameter regimes where long-range forces between atoms dominate, we use a truncated Taylor series expansion to derive the leading order behaviour of the kurtosis of the force distribution in the limit $n \to 0$. Finally, we utilise a Laplace integral approximation to ascertain the leading order behaviour of the standardised moments of force at low temperatures ($T \rightarrow 0$). Results from simple MD simulation are presented to provide evidence for the efficacy of these methods and underlying assumptions.
 
This is followed by Section~\ref{ManyBody}, where the natural idea that long range force calculations dictate asymptotic behaviour is extended from the 1D model to many-body systems of arbitrary size in three spatial dimensions. These systems exhibit the physical properties of standard MD simulations: \textit{i.e.} cubic geometry with periodic boundary conditions that employ the minimum image convention. In particular, we can analyze the system by performing calculations on a central cubic cell. In Section~\ref{NumSim}, MD results are displayed for many-body systems. We present the dependence of the standardised force moments on density, $n$, and temperature, $T$, and discuss the parameters and integrator schemes utilised in producing the results of MD simulations.

\section{\label{Notation}Notation}

\noindent
We consider a system of $N$ identical atoms interacting  via the Lennard-Jones 12-6 potential~\cite{Jones:1924}. This is a ubiquitous inter-atomic pairwise potential; here the potential between atoms labelled $i,j=1,2,\dots,N$  positioned at $\mathbf{q}_i,\mathbf{q}_j \in \mathbb{R}^{3}$ is given (in reduced units~\cite{Frenkel:1996}) by the expression 
\begin{equation}
\label{LJ}
    U_{\scriptsize{ij}}(r_{\scriptsize{ij}})= 4\left(\frac{1}{r_{\scriptsize{ij}}^{12}}-\frac{1}{r_{\scriptsize{ij}}^6}\right),
\end{equation}
where $r_{\scriptsize{ij}}=\left|\mathbf{q}_i - \mathbf{q}_j\right|$ is the distance between atoms. The Lennard-Jones potential (\ref{LJ}) between two atoms has a unique minima obtained at $r_{\scriptsize{ij}}=r_*=2^{1/6}$.

We employ the framework of statistical mechanics for this closed many-body system and describe atom $i=1,2,\dots,N$ by phase space coordinates $\{\mathbf{q}_i,\mathbf{p}_i\} \in \mathbb{R}^6$, were $\mathbf{p}_i$ denotes the momentum of the $i$-th atom.
We work in the canonical ensemble with temperature $T$; the partition function therefore becomes 
\begin{equation}
\label{partitionfunction}
\mathcal{Z}_N(T,V)
=
\frac{1}{h^{3N} \, N!}
\iint\limits_{\Omega_{\mathbf{q}} 
\times \Omega_{\mathbf{p}}}
\exp [-\beta H(\mathbf{q},\mathbf{p})]
\,\, \mbox{d}^3\mathbf{q} \,\,
\mbox{d}^3 \mathbf{p} \, ,
\end{equation}
where $V$ is the volume of our closed system,
and 
$
\mathbf{q}=
(\mathbf{q}_1,\mathbf{q}_2,\dots,
\mathbf{q}_N)^T
$ 
and 
$\mathbf{p}
=
(\mathbf{p}_1,\mathbf{p}_2,\dots,
\mathbf{p}_N)^T$ are vectors containing the positions and momenta of all atoms in the system. Our integration domain is given by  $\Omega_{\mathbf{q}} \times \Omega_{\mathbf{p}} \subset \mathbb{R}^{3N} \times \mathbb{R}^{3N}$. This denotes the phase space of our system. For systems of interest $\Omega_{\mathbf{p}} \equiv \mathbb{R}^{3N}$. The underlying geometry of the system (and principle simulation cell) is a cubic box of size $L>0$, therefore $\Omega_{\mathbf{q}} \equiv (-L/2,L/2]\times \dots \times (-L/2,L/2]$. The phase space volume elements in equation~(\ref{partitionfunction})
are denoted by 
\begin{equation}
\label{dqdpdef}
\mbox{d}^3\mathbf{q}
=
\prod \limits_{i=1}^N \,
\mbox{d}^3\mathbf{q}_i 
\qquad
\mbox{and}
\qquad
\mbox{d}^3\mathbf{p}
=
\prod \limits_{i=1}^N
\,
\mbox{d}^3\mathbf{p}_i.
\end{equation}
Throughout this work we make use of reduced units~\cite{Frenkel:1996}, utilising Argon parameters~\cite{Rowley:1975}. In particular, all instances of $T$ in this work can be translated back to SI units with the transformation $T \rightarrow k_B T$ where $k_B$ is the Boltzmann factor. Therefore, in the partition function~(\ref{partitionfunction}), we have $\beta=1/T$ and $h$ is the Planck constant ($\approx 0.186$ in reduced units).
Finally, $H(\mathbf{q},\mathbf{p})$
is the classical Hamiltonian 
$H(\mathbf{q},\mathbf{p})
=
K(\mathbf{p})+U(\mathbf{q})$ 
with kinetic energy $
K(\mathbf{p})=
|\mathbf{p}|^{2}/2
$ (where the usual factor of mass is unity under reduced units) and a general potential $U(\mathbf{q})$.
The statistical average of a quantity $X$ for this $N$-body system is given by 
\begin{equation}
\label{eq:thermoavg}
\langle X \rangle
=
\frac{1}{\mathcal{Z}_N \, h^{3N} \, N!
}\iint\limits_{\Omega_{\mathbf{q}} 
\times \Omega_{\mathbf{p}}}
\, X \, 
\exp[-\beta H(\mathbf{q},\mathbf{p})]
\,\, \mbox{d}^3\mathbf{q} \,\,
\mbox{d}^3 \mathbf{p} \, ,
\end{equation}
where the Boltzmann factor acts as a statistical weighting for a configuration $\{\mathbf{q},\mathbf{p}\} \in \mathbb{R}^{6N}$, normalised such that $\langle 1 \rangle = 1$.

We label atoms so that the first one is the
tagged atom. Denoting the force on the 
tagged atom produced from the $j$-th atom
by 
$\mathbf{F}_{\scriptsize{j}} 
=
[F_{j,1}, F_{j,2}, F_{j,3}] \in \mathbb{R}^3$,
for $j=2,3,\dots,N$,
the total force 
$\mathbf{F}=[F_1,F_2,F_3]$ 
on the tagged atom is 
\begin{equation*}
\mathbf{F}=
\sum\limits_{\substack{j=2}}^{N}\mathbf{F}_{\scriptsize{j}}.
\end{equation*}
We define
\begin{equation}
\label{momentdef}
f_k
=
\int_{\Omega_{\mathbf{q}}} 
\hskip 1mm \, 
\left(\sum\limits_{\substack{j=2}}^{N}F_{\scriptsize{j,1}}
\right)^{\!\!k}
\, 
\exp[-\beta \, U(\mathbf{q})]
\; \mbox{d}^3\mathbf{q} \,\end{equation}
for $k=0,1,2,\dots.$
Then we have
\begin{equation*}
\frac{f_k}{f_0}  
=
\left\langle 
\left(\sum\limits_{\substack{j=2}}^{N}F_{\scriptsize{j,1}}
\right)^{\!\!k} \, \right\rangle
=
\langle 
F_1^k
\rangle.
\end{equation*}
Then the $k$-th standardised moment~(\ref{statkurt})
is given by
\begin{equation}
\alpha_k
=
\frac{f_0^{\, k/2-1} f_k}{f_2^{k/2}} \,,
\label{generalkurt}
\end{equation}
where we are interested in cases $k=4,6,8,\dots$. 

In order to study how the force distribution depends on the physical parameters of interest it is useful to identify how changes in these parameters will manifest themselves in the system. Indeed, we choose to work in the canonical ensemble with a target temperature of $T$: this is accomplished with the use of a thermostat which is discussed further in Section~\ref{NumSim} and Appendix~\ref{appb}. It is more illuminating to see that if we have a system with a fixed number of free interacting atoms $N$ in a cubic box of side $L$; the (reduced) number density is given by $n=N/L^3$. Therefore the approach we employ in this paper to ascertain how values of standardised moments depend on number density, will be to keep the number of atoms fixed but vary the box width $L$ - this will manifest as a change in density~$n$. Similarly one could keep the volume of the cubic box the same and vary the number of atoms though this is a point of discussion in Section~\ref{NumSim}.

For the remainder of the paper we will study systems with different spatial dimensions. The size of the system varies by changing the number of particles $N$; we will use equation (\ref{generalkurt}) as a crucial initial point in each calculation. We will naturally proceed by investigating systems of increasing complexity; starting from a cartoon one-dimensional model and culminating to a general many-body system of arbitrary size in three spatial dimensions.

\section{\label{1DTheory} One atom in a potential well}

\noindent
We now go on to illustrate three approaches to obtain the dependence of the force distribution on parameters $n$ and $T$. It is useful to note that, as we are now working in one spatial dimension, density $n$ is proportional to $1/L$, {\it i.e.} we have $n \propto 1/L$.  We will consider a simple system in one spatial dimension consisting of two atoms interacting through the Lennard-Jones potential~(\ref{LJ}) in interval $[0,L]$ with periodic boundary conditions. One of the atoms is considered to be fixed at
position $q_{\scriptsize{\mbox{0}}}=L/2 \in [0,L]$ and the other atom is free to move, therefore, we have $N=1$ free atom. Its position is denoted $x \in [0,L]$. Therefore, the inter-atomic distance is 
$r
=
|x 
- 
q_{\scriptsize{\mbox{0}}}|.
$
Using our simplified one-dimensional set up, $F_1 = F$ and $\Omega_{\mathbf{q}}=(0,L)$, equation~(\ref{momentdef}) reduces to
\begin{equation}
f_k(L)
=
\int\limits_{0}^{L}
F^k(|x-q_0|)
\,
\exp[-\beta \, U(|x-q_0|)]
\, \mbox{d}x,
\label{1dmoment}
\end{equation}
which is the marginalised expected value of the $k$-th moment of force $F(x)=-{\mbox{d}U}/{\mbox{d}x}$, where we have dropped subscripts in the Lennard-Jones potential~(\ref{LJ}) and we write it as $U(z)=4 (z^{-12} - z^{-6})$. Utilising the symmetry of the potential (and therefore the force) we are left with 
\begin{equation}
\label{1Dgeneral}
f_k(L)
=
2 \int\limits_{0}^{L/2}
F^k(r)
\,
\exp[-\beta \, U(r)]
\, \mbox{d}r.
\end{equation}
In what follows, we will assume that we are in a regime where the box width $L$ satisfies $L \gg r_*$, where $r_* = 2^{1/6}$ minimizes the Lennard-Jones potential $U$. 

\subsection{\label{PDE} Differential equation for standardised moments}

\noindent
We consider a perturbation of the form $L\rightarrow{L+\delta L}$. Using equation~(\ref{1Dgeneral}) and considering terms to the order $O(\delta L)$, we obtain
\begin{eqnarray*}
&&f_k(L+\delta L) 
= 
f_k(L)
\, + \, 
f_k'(L) \; \delta L
\, + \,
O\big(\delta L^2\big) 
\\
&&
= 
f_k(L)
\, + \,
 F^k(L/2)
\,
\exp[-\beta \, U(L/2)]
\,
\delta L
\, +
\,
O\big(\delta L^2\big).
\end{eqnarray*}
Using equation~(\ref{generalkurt}),
we approximate $\alpha_k(L+\delta L)$ by
\begin{equation*}
\alpha_k(L) 
+
\alpha_k(L) 
\,
\upsilon_k(L) 
\,
\exp[-\beta \, U(L/2)]
\,
\delta L
+
O\big(\delta L^2\big),
\end{equation*}
where our notation $\alpha_k(L)$ highlights the dependence of the standardised moments of force, $\alpha_k$, on $L$, and function $\upsilon_k(L)$ is given by
\begin{equation}
\upsilon_k(L)
=
\frac{k-2}{2 \, f_0(L)}
+ \frac{F^k(L/2)}{f_k(L)} 
- 
\frac{ k \, F^2(L/2)}{2 \, f_2(L)}
\,.
\label{upkl}
\end{equation}
Taking the limit 
${\delta L \to 0}$, we obtain the derivative 
of the $k$-th standardised moment of force,
with respect to L, as
\begin{equation}
\label{partial1L}
\frac{\partial \alpha_k}{\partial L}
(L)
=
\upsilon_k(L) 
\,
\exp[-\beta \, U(L/2)]
\,
\alpha_k(L),
\end{equation}
where $\upsilon_k(L)$ are expressed in terms of integrals~(\ref{1Dgeneral}) as given by equation~(\ref{upkl}).

\subsection{\label{Integralapprox} Far-field integral approximation}

\noindent
To further analyze integrals~(\ref{1Dgeneral}), 
we introduce a cutoff $c$, which satisfies that
$r_*<c<L/2$, where $r_*=2^{1/6}$ is a unique 
maximum of $\exp[- \beta \, U(z)]$, which can 
be Taylor expanded as 
$
\beta
(1+4z^{-6}+4z^{-12}-16/3z^{-18}+8z^{-24} \dots)$.
Considering sufficiently large $L$, we can choose the cutoff $c$, so that
\begin{equation}
\left|
f_0(L)
- 
2 \left(
\int\limits_{0}^{c}
\exp[-\beta \, U(r)]
\, \mbox{d}r
+ 
\beta \!\!\!\!
\int\limits_{c}^{L/2}
\!\!
1+\frac{4}{r^6} \, \mbox{d}r \right)\right|\leq \varepsilon, \,
\label{eqtol}
\end{equation}
where tolerance $\varepsilon$
is chosen to be $10^{-4}$ in our illustrative computations.
This splitting allows us to numerically calculate the bulk of the integral~(\ref{1Dgeneral}) as a constant independent of $L$ and then use the second term to give an analytic expression for $\alpha_k$ with dependence on $L$, and ultimately on $n$.

The range of values of $T$ that are of typical use are chosen in order to maintain the liquid state of Argon during simulation. These are approximately temperatures in the interval $0.70 < T < 0.73$ under ambient conditions~\cite{Lide:2004}. Therefore, as volume is varied we are in a regime where $\beta=O(1)$, for convenience we set $\beta=1$. Though given that the density of our system changes between each simulation some systems will be in a liquid phase and others in a gaseous phase, this is a point of discussion in Section~\ref{NumSim}.

Splitting the integration domain $[0,L/2]$ of integral~(\ref{1Dgeneral}) into $[0,c]$ and $[c,L/2]$, 
we use the exact form of the integrand
in $[0,c]$ to obtain a `near-field' contribution. Utilising an approximate form for the integrand given by the truncated Taylor expansion $f(z)$ in the domain $[c,L/2]$ gives rise to a density dependent `far-field' contribution. Combining these we arrive at the approximate form for $f_0(L)$. Using cutoff~$c=2$, 
equation~(\ref{eqtol}) is
satisfied with $\varepsilon = 10^{-4}$. Therefore, upon numerically calculating the bulk contribution for the integral with domain $[0,2]$, we get
\begin{equation}
f_0(L)
=
2 \int\limits_{0}^{L/2}
\exp[-\beta \, U(r)]
\, \mbox{d}r 
\;=\; b_0 + L + O\!\left(L^{-6}\right)
\label{1dljforce0}
\end{equation}
with $b_0 =-0.71832$, which depends on our choice of cutoff $c=2$.
Similarly, we can calculate far-field integral approximations
of integrals~(\ref{1Dgeneral}) for general values of $k=2,4,6,8,10,12$.
The integrand $F^k(r)
\,
\exp[-\beta \, U(r)]$
has maxima when $r=r_*=2^{1/6}$ or when $k \, U''(r)= \beta \, (U'(r))^2$. 
This forms a cubic in $r^6$ that can be solved. For the values of $k$ used in this work, this sometimes results in a global maximum, that always lies at a distance less than $r<r_*$ from the origin. Therefore $r_*=2^{1/6}$ is the furthest maximum of the integrand from the origin. 

Splitting integral~(\ref{1Dgeneral}) into a near-field and far-field contribution, using the general cutoff $c=2$, we find
\begin{equation}
f_k(L)
\;=\; b_k + O\!\left(L^{-7k}\right)
\label{1dljforce}
\end{equation}
The near-field contributions, $b_k$, generally increase vastly if we increase the value of~$k$, for example 
\begin{equation}
b_0 =-0.71832,
\;\;
b_2=130.64 
\;\; \mbox{and} \;\; b_4=2.5727 \times 10^5, \;\;
\label{b0b2b4values}
\end{equation} while the dependence on $L$ decreases more rapidly for larger values of $k$. Therefore, the non-negligible density contributions to $\alpha_k(L)$ in the low density limit come exclusively from the normalisation $f_0(L)$ given by~(\ref{1dljforce0}).

\begin{figure}
{\leftskip -1.5mm
\includegraphics[width=9cm]{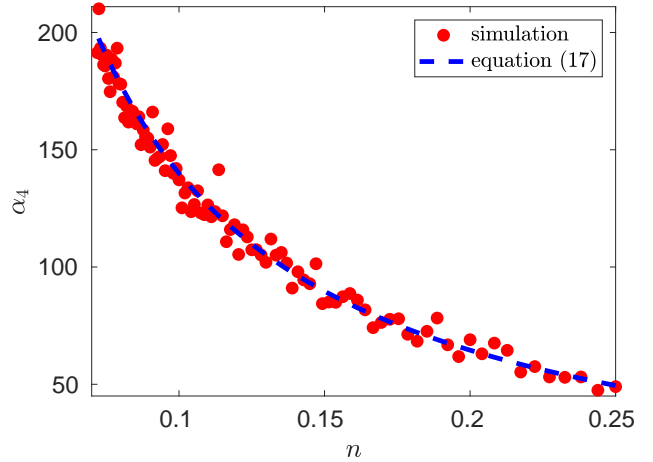}
\par}
\caption{{\it Plot of $\alpha_4$ as a function of $n=1/L$ for the illustrative one-atom system. Results of MD simulations are compared with $\alpha_4=-10.828+15.074 \, n^{-1}$ obtained by using equation~$(\ref{1dljkurtosisfinal})$ with $b_0$, $b_2$ and $b_4$ given by $(\ref{b0b2b4values})$ (blue dashed line).  MD simulation results for temperature $T=1$ utilising Langevin dynamics~\cite{Giro:1985} described in equation~{\rm (\ref{langevin})}, with friction parameter $\gamma=0.1$, are represented by red dots. The MD simulation length was a total of $1.1 \times 10^8$ time steps with the first $10^7$ time steps used for initialisation.}}
\label{1Da4}
\end{figure}

Substituting equations~(\ref{1dljforce0}) and (\ref{1dljforce}) in equation~(\ref{generalkurt}), we obtain an expression for the general $k$-th standardised moment of force
\begin{equation}
\alpha_k(L)
=
\frac{b_0^{k/2-1} b_k}{b_2^{k/2}} 
\left( 1+\frac{L}{b_0}+O\!\left(L^{-6}\right)\right)^{k/2-1}\!\!\!\!.
\label{1dljkurtosisfinal}
\end{equation}
Using the values of $b_0$, $b_2$ and $b_4$ given by (\ref{b0b2b4values}),
we obtain the dependence of the kurtosis of the force distribution on the reduced number density $n=1/L$ in the dilute limit~$n \to 0$ as
$
\alpha_4
=
- 10.828 
+ 
15.074 \, {n}^{-1}+O\big(n^6\big).
$
Figure~\ref{1Da4} compares this result with the results obtained by MD simulation of the one atom system. We observe that MD is in good agreement with the 
results obtained by formula~(\ref{1dljkurtosisfinal}).

\subsection{Leading order behaviour for differential equation~(\ref{partial1L})}

\begin{figure}
{\leftskip -1.5mm
\includegraphics[width=9cm]{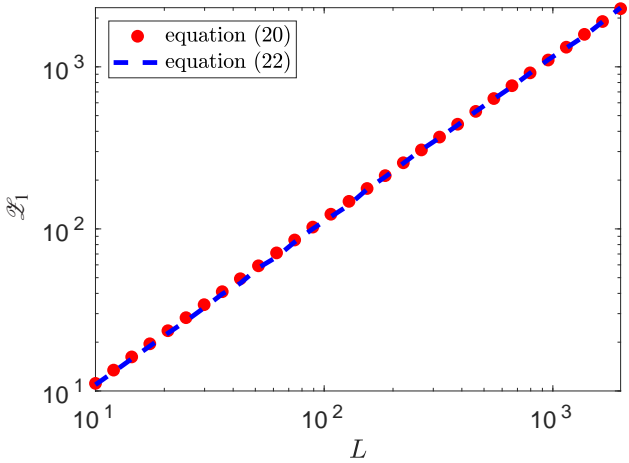}
\par
\vskip -3.7cm
\noindent \rotatebox{90}{$\mathcal{Z}_1$}
\vskip 3cm}
\caption{\textit{Approximation of the partition function $\mathcal{Z}_1(T,V)$ obtained using the right hand side of equation~$(\ref{partition1D})$ with $k=4$ and values of kurtosis ($\alpha_4$) estimated from MD simulation (blue dashed line). The exact values obtained by~$(\ref{defz1})$ 
are plotted as the red dots.}}
\label{1Dpartition}
\end{figure}

\noindent
Since $L/2 > r_*$, the force $F(L/2)$ monotonically decreases as a function of $L$.
When looking at leading order approximations in the low density limit $n \to 0$ (equivalent to limit $L \to \infty$) to equation (\ref{partial1L}), we need to analyse $\upsilon_k(L)$. The second and third term in equation (\ref{upkl}) converge to zero more rapidly than the first term as $L \to \infty$, therefore the leading order behaviour is given by the first term
\begin{equation}
\upsilon_k(L)
\sim 
\frac{k-2}{2 \, f_0(L)}
\qquad \hbox{as} \qquad  L \rightarrow{\infty}.
\label{lowdenlimuk}
\end{equation}
\noindent
By utilising the far field integral
approximation~(\ref{1dljforce0}),
we arrive at $f_0(L) \sim (b_0+ L)$, where $b_0=b_0(c)$ is a constant term that depends on cutoff parameter $c$. With this, our leading order approximation of the $k$-th standardised moment, $\alpha_k^0$, 
obeys 
\begin{equation*}
\frac{\partial \alpha_k^0}{\partial L}
(L)
=
\frac{k-2}{2 \, (b_0 + L)}
\,
\alpha_k^0(L) \, .
\end{equation*}
Finally this gives us that
\begin{equation}
\label{generallinearity1d}
\alpha_k^0(L)
=
C_k \, (b_0 + L)^{k/2-1}
=
C_k \left(
b_0+ {n}^{\,-1}
\right)^{k/2-1}
{\hskip -1mm},{\hskip 1mm}
\end{equation}
where $n=1/L$ is the reduced number density and $C_k$ is a
constant.
Equation~(\ref{generallinearity1d}) gives the same leading order
behaviour $n^{1-k/2}$ in the limit $n \to 0$ as equation~(\ref{1dljkurtosisfinal}): the same behaviour is also seen for the Lennard-Jones fluid in Section~\ref{ManyBody}. Though the method above is more generally applicable to include potentials that monotonically decay as $r^{-a}$ as $r \to \infty$ for $a>0$. 
We next make the observation that equation (\ref{partitionfunction}) in 1D can be written as:
\begin{equation}
\mathcal{Z}_1(T,V)
=
\frac{1}{h}
\int\limits_0^{L}
\exp [-\beta \, U(q)]
\,\, \mbox{d}q 
\int\limits_{-\infty}^{\infty} \!\!
\exp \! \left[-\frac{\beta \, p^2}{2} \right]
\,\, \mbox{d}p \, , \;
\label{defz1}
\end{equation}
where the Planck factor of $1/h$ arises instead of $1/h^3$ due to the fact that we are in one-dimensional physical space.
Using~(\ref{1Dgeneral}), we obtain
\begin{equation}
f_0(L)
=
h \, \mathcal{Z}_1(T,V) \, \sqrt{\frac{\beta}{2\pi}}.
\label{f0useful}
\end{equation}
Considering the low density limit $n \to 0$ (i.e. $L \to \infty$) in equation (\ref{partial1L}) and using~(\ref{lowdenlimuk}) and (\ref{f0useful}), we obtain
\begin{equation}
\label{partition1D}
\mathcal{Z}_1(T,V) 
\sim 
\frac{(k-2) \sqrt{2\pi}}{\sqrt{h^2 \, \beta}} \,  
\left[\alpha_k(L) \left(
\frac{\partial \alpha_k}{\partial L}(L)
\right)^{\!\!-1} \right],
\end{equation}
as $L \to \infty$. In particular, we can obtain the partition function~(\ref{defz1}) in the dilute
(low density) limit by using information about the moments of the force distribution. The accuracy of equation~(\ref{partition1D}) is illustrated in Figure~\ref{1Dpartition}, where we use $k=4$. We use
MD simulations of a single atom, using a range of simulation box widths $L$. We estimate the values of kurtosis of the force distribution, its derivative with respect of $L$ and use the right hand side of equation~(\ref{partition1D}) to estimate the $\mathcal{Z}_1(T,V)$. Considering $L \ge 10$, the result is within 5\% error when compared with the exact result~(\ref{defz1}), while for larger values of box width $L$ the error decreases to around 1\%, confirming that the
formula~(\ref{partition1D}) is valid in the asymptotic limit $L \to \infty$.

\subsection{\label{TempDep}
Temperature dependence of standardised moments}

\noindent
One can perform a similar analysis as in Section~\ref{PDE}, viewing the moments $\alpha_k=\alpha_k(T)$ as a function of temperature $T=1/\beta$. To do that, we consider the moment definition~(\ref{1Dgeneral}) as a function of temperature~$T$, namely, we define
\begin{equation}
\label{1DgeneralT}
f_k(T)
=
2 \int\limits_{0}^{L/2}
F^k(r)
\,
\exp \!\left[- \frac{U(r)}{T} \right]
\, \mbox{d}r.
\end{equation}
Considering small perturbations of these functions with respect to $T \rightarrow{T+\delta T}$, while fixing the domain length $L$, and collecting terms up to first order in $\delta T$, we obtain
\begin{equation}
\label{partial1T}
\frac{\partial \alpha_k}{\partial T}(T) 
= \nu_k(T) \, \alpha_k(T) \,, 
\end{equation}
where
\begin{equation}
\label{constants2}
\nu_k(T) 
= 
\left( \frac{k}{2}-1 \right)\frac{f'_0(T)}{f_0(T)}
+ \frac{f'_k(T)}{f_k(T)} - \left(\frac{k}{2}\right)\frac{f'_2(T)}{f_2(T)}\,.  
\end{equation}
Combining equations~(\ref{partial1T}) and (\ref{constants2}) with equation~(\ref{f0useful})
where $\beta=1/T$, we obtain 
\begin{equation*}
\frac{\partial}{\partial T} 
\ln \!
\left(
\frac{\alpha_k^2 f_2^k}{f_k^2} \right)
=\left( k - 2 \right)\left(\frac{\partial}{\partial T} \ln (\mathcal{Z}_1) -\frac{1}{2T} \right).
\end{equation*}
Since
$- \partial/ \partial \beta (\ln{\mathcal{Z}_1})$
is equal to the average energy of the system,
$\langle E \rangle$, we have 
\begin{equation}
\langle E \rangle
=
\frac{T}{2}
+
\frac{T^2}{k-2} \,
\frac{\partial}{\partial T} 
\ln \!
\left(
\frac{\alpha_k^2 f_2^k}{f_k^2} \right),
\label{avgenergyeqn}
\end{equation}
where the first term on the right hand side of equation~(\ref{avgenergyeqn}) is the average kinetic energy of our one-atom system. Substituting equation~(\ref{generalkurt})
into the second term on the right hand side, it can be rewritten as $T^2 \partial(\ln{f_0}) / \partial T$.
Thus, using equation~(\ref{eq:thermoavg}), we confirm that the second term on the right hand side of equation~(\ref{avgenergyeqn}) is the average potential energy.

\subsection{\label{Laplace} Low temperature limit}

\noindent 
Next, we consider the behaviour of the $k$-th standardised moment of force, $\alpha_k(T)$, given by equation (\ref{generalkurt}), in 
the low temperature limit, $T \to 0$, which is equivalent to the limit~$\beta \rightarrow{\infty}$. Since the inter-atomic potential $U(r)$ has a global minimum at $r=r_*$ in interval $[0,L/2]$, integrals of the form~(\ref{1Dgeneral}) and~(\ref{1DgeneralT}) can be approximated by Laplace's method in the limit $\beta \rightarrow{\infty}$ and $T \to 0$, respectively. A general discussion of Laplace's method is given in Chapter 6 of the book by Bender and Orszag~\cite{Bender:1999}. We calculate the asymptotic expansion of  $f_0(T)$ by
applying Laplace's method to integral~(\ref{1DgeneralT}) for $k=0$. We approximate the integration limits of integral~(\ref{1DgeneralT}) to lie within the domain $r \in (r_*-\varepsilon,r_*+\varepsilon)$, where $\varepsilon \ll 1$, and we Taylor expand $U(r)$ at $r=r_*$. Using $U'(r_*)=0$, we have
\begin{equation*}
\begin{split}
U(r) & \approx U(r_*)+(r-r_*)^2U''(r_*)/2 \\
& +(r-r_*)^3U^{(3)}(r_*)/6+(r-r_*)^4U^{(4)}(r_*)/24 \,,
\end{split}
\end{equation*}
where we denote the $m^{\hbox{\scriptsize{th}}}$ derivative of $U$ as $U^{(m)}$ for $m \ge 3$.
Substituting into integral~(\ref{1DgeneralT}), we arrive at the asymptotic expansion
\begin{equation}
f_0(T)
\sim 
\frac{\sqrt{\pi \, T}
\, 
\exp[-
U(r_*)/T]}{\sqrt{2\,
U''(r_*)}}\Big[1
+
B_0 \, T+O\big(T^2\big)\Big],
\label{1dtemplaplacef0}
\end{equation}
as $T \to 0$, where constant $B_0$ is given by~\cite{Bender:1999} 
\begin{equation}
B_0
=
\frac{5 \, (U^{(3)}(r_*))^2}
{
24 \, (U''(r_*))^3}- 
\frac{U^{(4)}(r_*)}{8 \, (U''(r_*))^2} \,.
\label{eqb0}
\end{equation}
To apply Laplace's method to integral~(\ref{1DgeneralT}) for $k=2,4,6,\dots$, we note that $F^k(r)=(U'(r))^k$ for even values of $k$. 
Using the truncated Taylor expansion around $r=r_*$
and noting that $U'(r_*) = 0$, we have
\begin{eqnarray}
F^k(r) 
\, \approx \, (r-r_*)^k \,
\Big(&&
\left(U''(r_*)\right)^{k} 
\, + \, 
(r-r_*) \, C_{k,1}
\nonumber
\\ && \, + \, 
(r-r_*)^{2} \, C_{k,2}
\Big),
\label{taylorck1ck2}
\end{eqnarray} 
where $C_{k,1}$ and $C_{k,2}$ are constants, which can be expressed in terms of the derivatives of potential $U(r)$ at
$r=r_*$ (see equations~(\ref{eqck1}) and (\ref{eqck2}) in Appendix~\ref{appck1ck2}).
This gives the asymptotic expansion
\begin{equation}
\begin{split}
f_k(T)
& \sim 
\frac{\sqrt{\pi \, T}
\, 
\exp[-
U(r_*)/T]}{\sqrt{2\,
U''(r_*)}}
\\
& \times 
\, A_{k}
\left[
\, T^{k/2}
\, + \,
B_{k}
\, T^{k/2+1}
+O\!\left(\!T^{k/2+2}\!\right)\right] .
    \label{1dtemplaplacegen}
    \end{split}
\end{equation}
as $T \to 0$, where constants $A_{k}$ and $B_{k}$ are given by
\begin{equation*}
A_{k}
=
\left(U''(r_*)\right)^{k/2}  (k-1)!!
\end{equation*}
and 
\begin{equation*}
B_{k} 
\, = \,
\frac{(4k-15) \, (k^2-1) (U^{(3)}(r_*))^2}{72 \, (U''(r_*))^{3}} 
+
\frac{(k^2-1) \, U^{(4)}(r_*)}{8 \, (U''(r_*))^{2}} \, ,
\end{equation*}
where the last formula reduces to equation~(\ref{eqb0}) for $k=0$.
Substituting (\ref{1dtemplaplacef0}) and (\ref{1dtemplaplacegen}) into (\ref{generalkurt}) gives the following expression in the limit $T \to 0$:
\begin{equation*}
\alpha_k 
\sim
(k-1)!! \,
\left(
1
\, +
\, \frac{ \left( k-2 \right) B_0 + 2 B_{k} - k \, B_{2}}{2} \; T \, + \, O\big(T^2\big)\right).
\end{equation*}
In particular, we have $\alpha_2 \sim 1 + O\big(T^2\big)$ and
\begin{eqnarray}
\alpha_4 
\, &\sim& \, 
3 \, + \, 
3 \left( 
\frac{(U^{(3)}(r_*))^2}{(U''(r_*))^{3}} 
+
\frac{U^{(4)}(r_*)}{(U''(r_*))^{2}} \right) T + O\big(T^2\big) \nonumber \\
&=&
3 +  \frac{203}{6} \, T + O\big(T^2\big).
\label{reslapl}
\end{eqnarray}
Therefore, Laplace's method predicts that the standardised moments of the force distribution, $\alpha_k(T)$, tend to the values given in equation (\ref{gaussmoments}) for Gaussian moments in the low temperature limit. This limiting behaviour is to be expected as during the Laplace approximation we use a Gaussian distribution to approximate the Boltzmann factor. We can interpret this approach as approximating the force distribution as Gaussian and perturbations of the system around small temperatures give rise to non-Gaussian contributions to the standardised moments. 

\begin{figure}
\includegraphics[width=9cm]{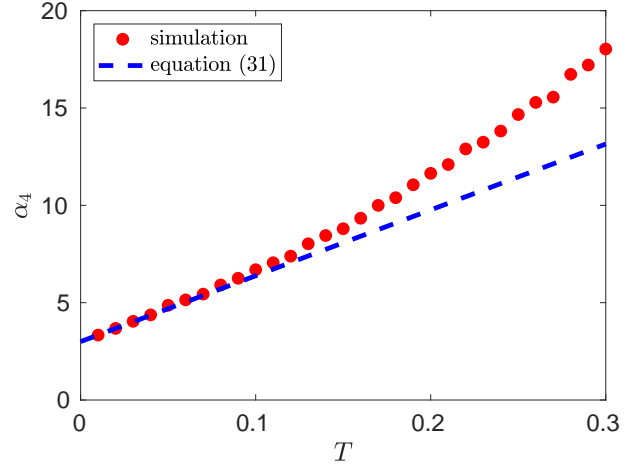}
\caption{{\it   
Kurtosis, $\alpha_4$, as a function of temperature, $T$, for $T \le 0.3$. The linear behaviour is estimated as $\alpha_4(T) \sim 2.9388+37.002 \, T$ for $T\in (0.01,0.10)$ (using the MD computed data, with density $n=0.1$, visualized as red dots). We compare this to the theoretical linear result $3 + 203 \, T/6$ predicted by equation~$(\ref{reslapl})$ (illustrated by the blue dashed line).
}}
\label{laplacetemp1d}
\end{figure}

Results from MD simulation are illustrated in Figure~\ref{laplacetemp1d} over the range of values of temperature $T$. We see that the behaviour of kurtosis, $\alpha_4$, is well approximated by the linear approximation $3 + 203 \, T/6$ given in equation~(\ref{reslapl}) for the temperature values satisfying $T \le 0.1$, though this agreement diverges as temperature $T$ increases and higher order terms, $O\big(T^2\big)$ in equation~(\ref{reslapl}), become significant. In Figure~\ref{laplacetemp1d}, we fix the box width as $L=10$. Increasing the box width much further would take us to a regime where the particle is essentially free and the approximation calculated by the Laplace method around the potential minimum would lose validity.

\section{\label{ManyBody} Many-body systems}

\noindent
In this section we employ the far field approximation approach introduced in Section~\ref{Integralapprox} and we will vary the number density of the system by changing the size $L$ of the integration domain, which will be given as the three-dimensional cube $[0,L]^3.$ Using notation introduced in Section~\ref{Notation}, the distance between atoms labelled $i,j=1,2,\dots,N$  positioned at $\mathbf{q}_i,\mathbf{q}_j \in \mathbb{R}^{3}$ is denoted by $r_{\scriptsize{ij}}=\left|\mathbf{q}_i - \mathbf{q}_j\right|$. Taking into account the periodic boundary conditions, the distance 
$\left|\mathbf{q}_i - \mathbf{q}_j\right|$ is
the minimum image inter-atomic distance given by
\begin{equation}
|\mathbf{q}_i-\mathbf{q}_j|
=
\left(\overline{(q_i^x-q_j^x)}^2+\overline{(q_i^y-q_j^y)}^2+\overline{(q_i^z-q_j^z)}^2 \right)^{1/2} \!\!\!\!,
\label{mimdist}
\end{equation}
where the overline denotes $\overline{\zeta} = \zeta - L \, [\zeta/L]$ for $\zeta \in \mathbb{R}$ and $[.]$ rounds a real number to the nearest integer. For an  interacting $N$-body system the dimensionality of the integral given by equation (\ref{momentdef}) is $3N$. We first present an illustrative calculation with $N=2$ interacting atoms in Section~\ref{secNis2}
and then we study systems with larger values of $N$ in Section~\ref{NumSim}. 

\subsection{\label{secNis2} Dependence of $\alpha_k$ on density for $N=2$ interacting atoms}

\noindent
In Section~\ref{1DTheory}, we have considered two atoms in the one-dimensional spatial domain, where one atom was fixed at position $q_0,$ i.e. we have effectively studied a single atom in a one-dimensional potential well. Here, we will consider $N=2$ interacting atoms in the three-dimensional cubic domain $[0,L]^3$ with periodic boundary conditions. We calculate the $k$-th standardised moment of force according to equation~(\ref{generalkurt}). To do so, we consider equation~(\ref{momentdef}), where we have  
$
\mbox{d}^3\mathbf{q}
=
\mbox{d}^3\mathbf{q}_1
\,
\mbox{d}^3\mathbf{q}_2,
$
$
U(\mathbf{q})
=
U(r_{\scriptsize{12}}),
$
$
F_1(\mathbf{q})
=
F_1(r_{\scriptsize{12}})
$
and we integrate over the domain $\Omega = [0,L]^3 \times [0,L]^3$ to get
\begin{equation}
\label{f0N2}
f_k
=
\int_{\Omega} \, 
F_1^k(r_{\scriptsize{12}})
\,
\exp[-\beta \, U(r_{\scriptsize{12}})]
\; 
\mbox{d}^3\mathbf{q}_1
\,
\mbox{d}^3\mathbf{q}_2. \end{equation}
It is useful to introduce a change of coordinates $\xi^\ell=q_1^\ell-q_2^\ell$ and $\eta^\ell=q_1^\ell+q_2^\ell$ for $\ell=x,y,z$. We note that $r_{12}$ is only dependent on the $\xi^\ell$ variables, therefore one can trivially integrate~(\ref{f0N2}) through the $\eta^\ell$ variables as the integrand has no dependence on these to obtain
\begin{equation*}
f_k 
= 
\frac{L^3}{8}
\int\limits_{-L}^{L} \int\limits_{-L}^{L} \int\limits_{-L}^{L}  \, 
F_1^k(r_{\scriptsize{12}})
\,
\exp[-\beta \, U(r_{\scriptsize{12}})]
\;
\mbox{d}\xi^x  \, 
\mbox{d}\xi^y  \, 
\mbox{d}\xi^z  \, ,
\end{equation*}
where $r_{\scriptsize{12}}$ is the minimum image inter-atomic distance~(\ref{mimdist}).
This integral can be written in terms of standard Euclidean distance  $r^2=(\xi^x)^2+(\xi^y)^2+(\xi^z)^2$ as
\begin{equation}
\label{f0inta}
f_k 
= 
8 \, L^3
\int\limits_{0}^{L/2}
\;
\int\limits_{0}^{L/2} 
\;
\int\limits_{0}^{L/2} \,
F_1^k(r)
\,
\exp[-\beta \, U(r)]
\;
\mbox{d}{\boldsymbol \xi} \, ,
\end{equation}
where $\mbox{d}{\boldsymbol \xi}
= \mbox{d}\xi^x  \, 
\mbox{d}\xi^y  \, 
\mbox{d}\xi^z$. In order to analyse $f_k$ further by implementing a far field approximation, we need to make sure we are in a regime where the integrand is small - we do this by introducing a cutoff $\gamma$,
which will divide the cube $[0,L/2]^3$
into 8 cuboid subdomains, including
\begin{eqnarray*}
&&\Omega_1 = [0,\gamma]^3, 
\hskip 2.02cm
\Omega_2 = [0,\gamma]^2 \times 
[\gamma,L/2], \\
&&\Omega_3 = [0,\gamma] \times [\gamma,L/2]^2,
\hskip 6mm
\Omega_4 = [\gamma,L/2]^3.
\end{eqnarray*}
Utilising the symmetry of the problem, we can rewrite integral~(\ref{f0inta}) as
\begin{equation}
f_k
= 
8 \, L^3
\raise -0.3mm
\hbox{$\Bigg($}
\raise 0.7mm
\hbox{$\displaystyle
\;
\int\limits_{\Omega_1}$}
+
\;
3
\raise 0.7mm
\hbox{$\displaystyle
\int\limits_{\Omega_2}$}
+
\;
3
\raise 0.7mm
\hbox{$\displaystyle
\int\limits_{\Omega_3}$}
+
\raise 0.7mm
\hbox{$\displaystyle
\int\limits_{\Omega_4}$}
\raise -0.3mm
\hbox{$\Bigg)$}
\,
F_1^k(r)
\,
\exp[-\beta \, U(r)]
\;
\mbox{d}{\boldsymbol \xi}  \, .
\label{fkfourint}
\end{equation}
Considering (\ref{fkfourint}) for $k=0$, the integral over $\Omega_1$ is independent of $L$ and provides a bulk contribution to $f_0$ that will depend on~$\gamma$. The remaining three terms have integration domains that allow the integrand to be accurately described by a Taylor expansion giving the leading order contribution in the asymptotic limit $L\to \infty$ as $f_0 \propto L^6$, which can be rewritten in terms of the density, $n$, in the form
\begin{equation}
\label{f02}
f_0 
\, \propto \,
n^{-2} \hskip5mm 
\mbox{as} \hskip3mm n \to 0.
\end{equation}
Considering $f_k$ for $k \neq 0$, the integral over $\Omega_1$ in equation~(\ref{fkfourint}) is again independent of $L$. However in the far field expansion the integrals over $\Omega_2$, $\Omega_3$ and $\Omega_4$ all decay with $L$ due to the force factor. As the integration domain has essentially been transformed into that of inter-atomic distances about the three coordinates, when we increase the domain length, the inter-atomic force necessarily decays to 0. Therefore in the limit $L \to \infty$ the dominant term arises from integrating over $\Omega_1$, and we see that, for $k = 2,4,6,8,\dots,$ 
\begin{equation}
\label{fk2}
f_k 
\, \propto \,
n^{-1} \hskip5mm 
\mbox{as} \hskip3mm n \to 0 \, .
\end{equation}
This leaves us with the final result that in the low density limit $n \to 0$, combining equation (\ref{generalkurt}) with asymptotic expressions~(\ref{f02}) and~(\ref{fk2}),
\begin{equation}
\alpha_k 
\, \propto \,
n^{1-k/2} \hskip5mm 
\mbox{as} \hskip3mm n \to 0 \, . 
\label{mbfinal}
\end{equation}
While this result has been calculated for $N=2$ interacting atoms, it is also confirmed for larger values of $N$ by estimating the k-th standardised moments using MD simulations, as it is shown in the next section.

\subsection{\label{NumSim} MD simulations with $N$ interacting atoms}

\noindent
In this section we present the results from MD simulations of many-body systems in three spatial dimensions using different values of $N$, including the case $N=2$ (analyzed in Section~\ref{secNis2}). Atoms are subject to pairwise interactions governed by a Lennard-Jones potential, given in equation (\ref{LJ}). For each system we use a velocity-Verlet~\cite{Verlet:1967} integrator and maintain the system in the canonical ensemble by incorporating a Nosé-Hoover thermostat~\cite{Nose:1984}, see Appendix~\ref{appb}. We perform two types of MD simulation studies: those that are used for studying how the number density, $n$, of a system affects standardised moments, and those that aim to probe temperature dependency. In all cases we utilise a time step $\Delta t=0.01$. In the case of the simulation with $N=2$ atoms, we initialise the positions of atoms by setting $\mathbf{q}_1=\mathbf{0}$ and $\mathbf{q}_2=(L/2,L/2,L/2)$,
whereas for the $N=8,\,64,\,512$ atom systems, we choose to initialise these on a uniform cubic lattice. 

\begin{table}
\begin{ruledtabular}
\begin{tabular}{lccr}
$N$ & $t_{\scriptsize{\mbox{sim}}}$ & $L_0$ & $n_0$\\
\hline
2 & $10^{9}$\rule{0pt}{4mm} & 5 & 0.016\\
8 & $10^7$ & 3 & 1/64\\
64 & $10^6$ & 5 & 1/64 \\
512 & $10^4$ & 10 & 1/64\\
\end{tabular}
\caption{\label{simparameters} {\it The length of MD simulation, $t_{\scriptsize{\mbox{{\rm sim}}}}$, the (smallest) box width, $L_0$, used for simulations with $N$ atoms and density $n_0$ for MD simulations with varying temperatures.}}
\end{ruledtabular}
\end{table}

The MD simulation parameters are summarised in Table~\ref{simparameters}, where $t_{\scriptsize{\mbox{sim}}}$ is the total simulation time used for calculating the required statistics, which is preceded by the initial simulation of length $t_{\scriptsize{\mbox{sim}}}/10$ used for equilibrating the system.  When investigating the number density dependence, we perform 20~simulations each with a box width of $L=L_0\times (6/5)^{i-1},$ where $i=1,2,\dots,20$ labels the simulation number and $L_0$ is the smallest cubic box width. We simulate the $N=8,64,512$-atom systems with $L_0=\,3,\,5,\,10$, respectively. This enables direct comparison because we can identify triplets of simulated systems corresponding to systems of the same number densities. The two-atom system however is simulated in a sparser regime with $L_0=5$. We calculate statistics on the fly for every time step, for every atom and for each coordinate - therefore we average the computed results over the number of time steps ($t_{\scriptsize{\mbox{sim}}}/\Delta t$) and atom coordinates ($3N$). In particular, the statistics are calculated over $3 \, N \, t_{\scriptsize{\mbox{sim}}}/\Delta t$ data points. This is equal to $6 \times 10^{11}$ (resp. $1.536 \times 10^9$) data points in the simulation with $N=2$ (resp. $N=512$) atoms.

\begin{figure*}
\includegraphics[width=\textwidth]{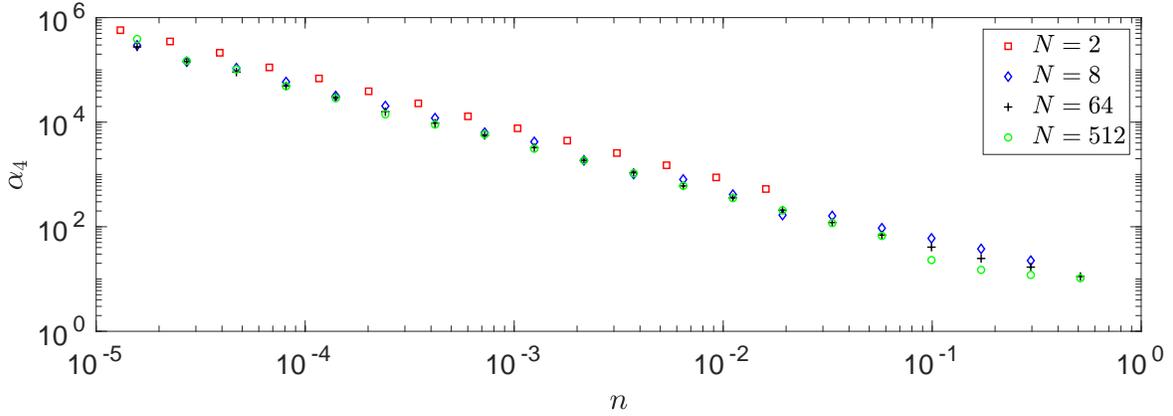}
\caption{{\it Dependence of kurtosis $\alpha_4$ on density $n.$ Each of the larger atomic systems ($N=8,64,512$) is simulated over the same domain of number densities, while the $N=2$ system is simulated in a sparser domain, though all are simulated in three spatial dimensions.  We truncate the results of the $N=2$ simulation in the plot, however the additional data points are used to calculate the results displayed in Figure~$\ref{mplengthfinal}$}.}
\label{ncomparison2}
\end{figure*}

Calculating the number density in three spatial dimensions by $n = N/L^3$, we can study the behaviour of kurtosis $\alpha_4$ as $n$ varies.
The results are presented in Figure~\ref{ncomparison2}. We see general agreement between behaviour of each of the four systems.
We see when $n$ is equal, the values of kurtosis are larger for $N=2$ than for the many-body systems with $N=8,64,512$, which agree well amongst themselves.

\begin{figure}
\includegraphics[width=9cm]{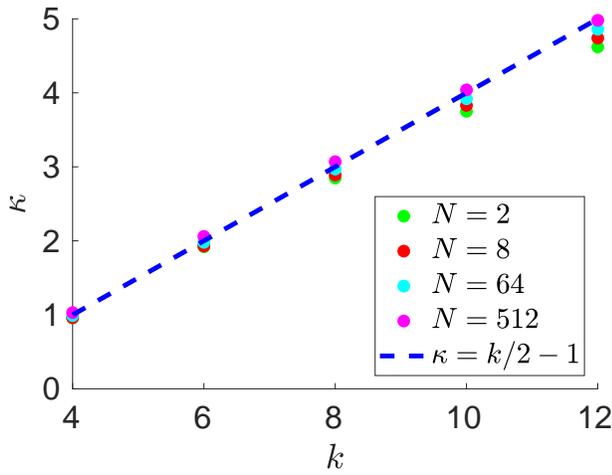}
\caption{{\it Comparison of the results of MD simulations for a range of values of the number of atoms, $N$. After long time simulation, we compute the asymptotic behaviour $\alpha_k \propto n^{-\kappa}$ and compare the leading order power scalings for each system. We compare this with the theoretical result~$(\ref{mbfinal})$ (denoted as a blue dashed line) that in the limit $n \to 0$ we expect the universal behaviour $\kappa=k/2-1$, where $k=2,4,6,\dots$ denotes which standardised moment of force we are looking at.}}
\label{mplengthfinal}
\end{figure}

The results in Figure~\ref{ncomparison2}
enable us to test the asymptotic expression~(\ref{mbfinal})
for $k=4$ derived in the limit $n \to 0.$ 
Utilising similar log-log plots for MD data, we estimate the power law behaviour of each standardised moment, $\alpha_k$, for $k=4,6,8,10,12$. Figure \ref{mplengthfinal} illustrates the results. 
All systems agree well with the predicted asymptotic behaviour~(\ref{mbfinal}), in particular the $N=512$ atom system. There is a slight deviation between the results due to the fact that the smaller atom systems require a larger $t_{\scriptsize{\mbox{sim}}}$ in order to converge fully to the predicted value. This discrepancy is amplified when looking at higher standardised moments due to the fact that we are calculating statistics resulting from $F_1^{12}$ (\textit{i.e.} for $\alpha_{12}$) compared to $F_1^4$ (\textit{i.e.} for $\alpha_{4}$), for example.

\begin{figure}
{\leftskip -1.5mm
\includegraphics[width=9cm]{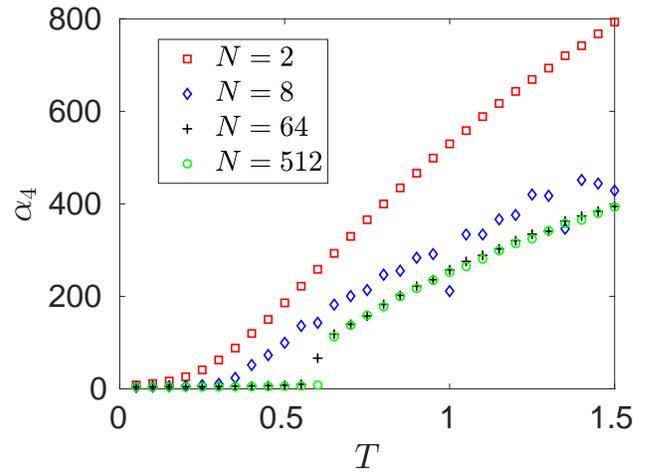}
\par}
\caption{{\it Dependence of kurtosis $\alpha_4$ on temperature $T$. Each atomic system is simulated at approximately the same density $n=n_0$ given in Table}~\ref{simparameters}.}
\label{tcomparison2}
\end{figure}

The dependence of kurtosis $\alpha_4$ on temperature $T$ is presented in Figure~\ref{tcomparison2}, where we keep the density fixed at $n = n_0$ given in Table~\ref{simparameters}. We observe that as temperature increases so does the kurtosis of the force distribution associated with each system. This can be explained in terms of the dynamics of the interacting atom system. If we maintain each system in the canonical ensemble, we expect on average that each atom will have a kinetic energy equivalent to $3 \, T/2$ (when in reduced units). As we increase this target temperature, the atoms become more energetic and thus are able to probe closer inter-atomic distances before a large repulsive force overcomes this inertial attraction. The range of forces on the tagged particle widens as temperature increases and therefore contributes to more outlier results in the distribution - leading to heavier tails and therefore distributions which become increasingly leptokurtic. 

\begin{figure}[htp]
\leftline{\hskip 1mm (a) \hskip 4.01cm (b)}
\noindent
\includegraphics[width=4.06cm]{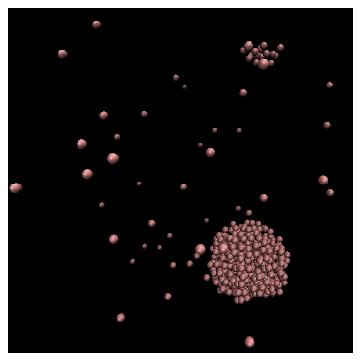} \;\; \includegraphics[width=4cm]{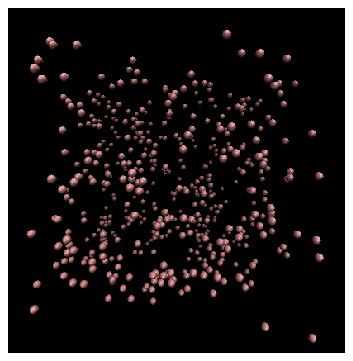}
\caption{{\it Snapshots~\cite{Humphrey:1996} of the MD simulation are taken for the system with $N=512$ atoms at time $t=7.5 \times 10^5$ for: {\rm (a)}
$T=0.6<T_*$; and {\rm (b)} $T=0.66>T_*$.
Density is $n=1/64$.
}}
\label{clustershots}
\end{figure}

In Figure~\ref{tcomparison2}, we observe that there is a qualitative difference between the results for $N=2$ and larger atom systems. 
We see a bifurcation for the $N=64$ and $N=512$ systems at some temperature $T_* \in (0.6,0.65)$, where a steady increase in kurtosis changes to a rapid increase. This bifurcation point in the phase plane lies on the coexistence boundary with $(n,T)=(1/64,T_*)$ and is due to a clustering mechanism which has been seen in MD simulations of Lennard-Jones fluids~\cite{Yoshii:1998}. From our results we see that the $N=2$ system has missed this behaviour completely.
Snapshots of the $N=512$-atom system at some $T=0.6 < T_*$, and $T=0.66>T_*$ are displayed in Figure~\ref{clustershots}. For $T=0.6$, we see a large cluster has formed in the many-atom system. There would be far fewer outlier force results in this case due to the fact that the large majority of atoms are moving as a collective and effectively have fixed inter-atomic forces. Compared to the $T=0.66$ snapshot, where we see that the atoms are too kinetically unstable to form these larger stable cluster structures, this results in more outlier forces felt between atoms due to the fact that the system is intrinsically more disordered. It is useful to note that this bifurcation point is located on the vapour-liquid coexistence boundary, the mechanisms of which have been studied on dilute Lennard-Jones fluids~\cite{Jung:2015}; here we see that this results in a bifurcation on standardised moments of the force distribution.

\begin{figure}
\includegraphics[width=9cm]{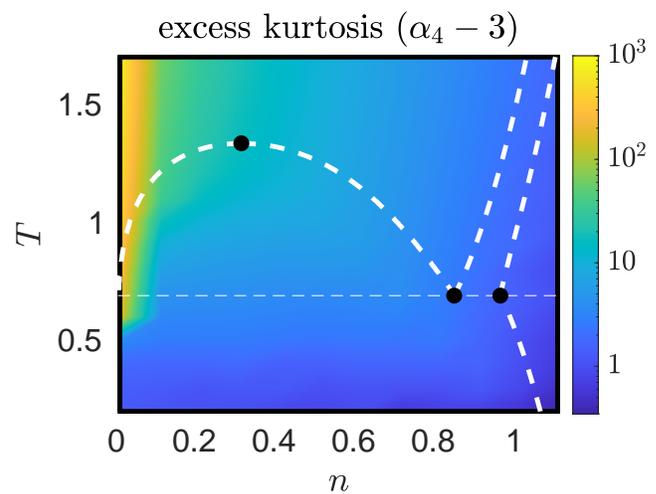}
\caption{{\it The excess kurtosis, $\alpha_4-3,$ calculated as a function of density $n$ and temperature $T$ for $n \le 1.11$ and $T \le 1.7$. The white dotted lines describe coexistence lines of different phases of a Lennard-Jones fluid taken from the literature~\cite{Schultz:2020,Stephan:2019:TPL,Stephan:2020:RCE,Mastny:2007}. The solid black dots indicate (from left to right), the critical point and vapour-liquid-solid triple points.}} 
\label{phaseplane}
\end{figure}

To understand the underlying variations of kurtosis, $\alpha_4$, with respect to changes in temperature and density, we use $12 \times 16$ MD simulations with $N=512$ atoms and $t_{\scriptsize{\mbox{sim}}}=3 \times 10^6$, varying simulation parameters $(n,T)$, where $n=10^{-2} + (i-1)/10$, for $i=1,2,\dots,12$, and $T=10^{-1}+j/10$, for $j=1,2,\dots,16$. The sampled values of excess kurtosis ($\alpha_4-3$) are displayed in Figure~\ref{phaseplane}. Here a bifurcation can be seen when using the smallest density $n=0.01$, as the change in colour is prominent in this vertical strip, indicating a large change of kurtosis. This occurs around $T=0.6$, which is consistent with the result in Figure~\ref{tcomparison2}, where we saw the bifurcation similarly located, though the slight shift in temperature is accounted for by the shift in density parameters used in each simulation (namely $n=0.01$ in Figure~\ref{phaseplane} and $n=1/64$ in Figure~\ref{tcomparison2}). 

In general, this low density strip contains the largest values of kurtosis, and covers much of the purely gas phase of the Lennard-Jones fluid. This paper has so far probed the low density limit in an attempt to understand why the standardised moments of force are so large, though Figure~\ref{phaseplane} gives a good overview that in general, regardless of phase, a decrease in temperature, or an increase in density, systematically lead to a lower value of standardised moments. In this case as $n \rightarrow \infty$ or $T \rightarrow 0$, we expect the $\alpha_4 \rightarrow 3$ (excess kurtosis tends to zero). This limiting regime corresponds to the solid phase of a Lennard-Jones system, where the force variations are minimal and the distribution is Gaussian. There is not enough space, nor energy, that lead to (many) outlier forces experienced by any atom, so the force distribution becomes less and less skewed from Gaussian, the deeper we probe in these regions. This intuition was demonstrated analytically in Section~\ref{Laplace} when we showed this limiting behaviour on a 1D cartoon model with equation~(\ref{reslapl}). 
It is interesting to note that these changes in values of $\alpha_4$ appear smooth about changes in temperature and density (in absence of the bifurcation point for larger values of $n$), regardless of phase transitions.

\section{\label{Conclusion} Discussion and conclusions}

\noindent
In Section~\ref{1DTheory} we have demonstrated use of a variety of methods to study the standardised moments of the force distribution in order to probe both their temperature and number density dependence. This gave way to a rich structure where we show that the partition function for a 1D system can be calculated entirely from these standardised moments. Extending the far field method introduced in Section~\ref{Integralapprox} to a system with $N$ atoms in three-dimensional physical space, Section~\ref{ManyBody} studies the dependence of $\alpha_k$ on number density $n$, deriving the asymptotic expression~(\ref{mbfinal}). Our analytic results are contrasted with MD simulations of four systems of $N=2,8,64,512$ interacting Lennard-Jones atoms and these are compared. 
The results agree well with theoretical predictions though the results for systems with larger values of $N$ are seen to converge more readily to the theoretically predicted results. In particular, rich dynamics such as clustering of Lennard-Jones fluids is completely missed by the systems with smaller values of $N$, but captured for systems with $N$ as small as $N=64$ atoms. In general, as temperature increases $\alpha_k$ increases due to energetic nature of atoms allowing them to push closer together and experience larger forces.  Clustering exhibited at the vapour-liquid coexistence phase incurs a bifurcation point whereby a large increase is seen in the standardised moments of force in Figure~\ref{tcomparison2}, though a general increase in temperature, or decrease in number density, results in an increase in $a_4$ regardless of the temperature/number density domain studied, as shown in Figure~\ref{phaseplane}.

\begin{acknowledgments}

\vskip -3mm

\noindent
This work was supported by the Royal Society [grant number RGF\textbackslash EA\textbackslash 180058] and by the Engineering and Physical Sciences Research Council [grant number EP/V047469/1].
\end{acknowledgments}

\appendix

\section{\label{appck1ck2}Constants $C_{k,1}$ and $C_{k,2}$ in equation~(\ref{taylorck1ck2})}

\vskip -3mm

\noindent
The constants appearing in equation (\ref{taylorck1ck2}), namely $C_{k,1}$ and $C_{k,2}$, are given by formulas
\begin{eqnarray}
C_{k,1}
& = &
\displaystyle
\frac{k \, (U''(r_*))^{k-1}  U^{(3)}(r_*)}{2} \, ,
\label{eqck1} \\ 
C_{k,2} 
& = & 
\displaystyle
\frac{k}{24}
\,
(U''(r_*))^{k-2} 
\label{eqck2} \\
&& \times
\bigg(
3 (k-1) 
\left(U^{(3)}(r_*)\right)^2 
\, + \,
4 \,
U''(r_*) \, U^{(4)}(r_*)
\bigg)
\, ,
\nonumber
\end{eqnarray} 
which can be derived in the following manner. Using $F^k(r)=(U'(r))^k$ for even values of $k$ and $U'(r_*) = 0$, we first note that 
\begin{eqnarray*}
F^{k,\,k}(r_*) 
&=&
k! \left(U''(r_*)\right)^{k},
\\
F^{k,\,m}(r_*) &=& 0,
\qquad\quad \mbox{for} \quad m \le k-1,
\end{eqnarray*}
where $F^{k,\,m}$ denotes the $m$-th derivative of $F^k$, i.e. the $m$-th derivative of the $k$-th power of $F$. Therefore, the first three non-zero terms of the Taylor expansion of $F^k(r)$ around $r=r_*$ are
\begin{eqnarray}
F^k(r) 
&\approx& (r-r_*)^k 
\left(U''(r_*)\right)^{k}
+ 
(r-r_*)^{k+1} \, \frac{F^{k,\, (k+1)}(r_*)}{(k+1)!}
\nonumber
\\
&&+(r-r_*)^{k+2} \, \frac{F^{k, \, (k+2)}(r_*)}{(k+2)!} \,.\label{taylorF}
\end{eqnarray} 
Therefore, we have
$C_{k,1}=F^{k,\, (k+1)}(r_*)/(k+1)!$
and $C_{k,2}=F^{k, \, (k+2)}(r_*)/(k+2)!$
and, to derive equations~(\ref{eqck1}) and (\ref{eqck2}), we need to express derivatives
$F^{k,\, m}(r_*)$ for $m=k+1$
and $m=k+2$ in terms of derivatives of $U(r)$ at $r=r_*.$
Using the product rule, the $m$-th derivative of $F^k(r)$ can be, in general, written as a finite sum of the form

\vskip -4.5mm

\begin{equation}
F^{k, \,m}(r)
=
\!\!\!\sum\limits_{\alpha_0,\alpha_1,\dots,\alpha_m=0}^k
\!\!\!\!\!\!
C(\alpha_0,\alpha_1,\dots,\alpha_m)
\,
\prod_{i=0}^m
\left(F^{(i)}(r)\right)^{\alpha_i} \!\!\!, \;\;
\label{termexp}
\end{equation}

\vskip -0.5mm

\noindent
where $F^{(i)}(r)$ is the $i$-th derivative of function $F(r)$ and $C(\alpha_0,\alpha_1,\dots,\alpha_m)$
are constants, many of them equal to zero. In fact, all terms in the expansion~(\ref{termexp}) have multiplicities that sum to $k$, that is we can only sum over sequences satisfying

\vskip -4mm
    
\begin{equation}
\sum\limits_{i=0}^{m} \alpha_i 
= k,
\label{expansion1}
\end{equation}

\vskip -0.5mm

\noindent
and all terms in the expansion~(\ref{termexp})  
have $m$ derivatives, that is,
we have

\vskip -7mm

\begin{equation}
\sum\limits_{i=0}^{m} i \, \alpha_i = m,
\label{expansion2}
\end{equation}

\vskip -0.5mm

\noindent
where $\alpha_i \in \{0,1,\dots,k\}$ for $i=0,1,2,\dots,m$.
Equation (\ref{expansion2}) is of the form of a finite Diophantine equation, which has no closed form for the number of solutions. In particular, simplifying equation~(\ref{termexp}) by solving equations~(\ref{expansion1})--(\ref{expansion2}) is, in general, not possible. However, noting the specific property that $F(r_*)=0=U^\prime(r_*),$ we see that all terms that have $\alpha_0 \neq 0$ will vanish when evaluated at this unique minimum $r=r_*$. In particular, we will obtain relatively simple forms of the sum~(\ref{termexp}) for $m=k+1$ and $m=k+2$ by considering equations (\ref{expansion1})--(\ref{expansion2}) with $\alpha_0=0$.

First, let us consider that $m=k+1.$ Using $\alpha_0=0$, there is only one solution of equations~(\ref{expansion1})--(\ref{expansion2}) in non-negative integers, namely $\alpha_1=k-1$,
$\alpha_2=1$ and
$\alpha_3=\alpha_4=\dots=0.$
Therefore, equation~(\ref{termexp}) implies

\vskip -4.5mm

\begin{equation*}
F^{k, \,(k+1)}(r_*)
=
C(0,k-1,1,0,\dots,0)
\,
\left(F^{(1)}(r_*)\right)^{k-1} 
\, F^{(2)}(r_*).
\end{equation*}

\vskip -0.5mm

\noindent
Using the general Leibniz rule~\cite{Traheem:2003:CNG}, we evaluate the combinatorial prefactor as
$C(0,k-1,1,0,\dots,0)
=
k (k+1)!/2.$ Substituting into $C_{k,1}=F^{k,\, (k+1)}(r_*)/(k+1)!$ and using $F(r_*)=-U^\prime(r_*)$ and that $k$ is an even integer, we obtain formula~(\ref{eqck1}).

Second, we consider the case $m=k+2.$ Using $\alpha_0=0$, there are two solutions of equations~(\ref{expansion1})--(\ref{expansion2}) in non-negative integers. The first solution is $\alpha_1=k-1$, $\alpha_2=0,$ 
$\alpha_3=1$ and
$\alpha_4=\alpha_5=\dots=0.$The second solution is  $\alpha_1=k-2$, $\alpha_2=2$
and $\alpha_3=\alpha_4=\dots=0.$ Therefore, equation~(\ref{termexp}) implies
\begin{eqnarray*}
&& F^{k, \,(k+2)}(r_*)
\\
&& \quad = \,
C(0,k-1,0,1,0,\dots,0)
\,
\left(F^{(1)}(r_*)\right)^{k-1}
F^{(3)}(r_*)
\\
&& \quad + \,
C(0,k-2,2,0,0,\dots,0)
\,
\left(F^{(1)}(r_*)\right)^{k-2}
\left(F^{(2)}(r_*)\right)^{2} \!\!\!.
\end{eqnarray*}
Using the general Leibniz rule~\cite{Traheem:2003:CNG}, we evaluate these combinatorial prefactors as
\begin{eqnarray*}
C(0,k-1,0,1,0,\dots,0)
&=&
\displaystyle
\frac{k }{6}
\,
(k+2)!,
\\
C(0,k-2,2,0,0,\dots,0)
&=&
\displaystyle
\frac{k (k-1) }{8} \, (k+2)!.
\end{eqnarray*}
Substituting into formula
$C_{k,2}=F^{k,\, (k+2)}(r_*)/(k+2)!$ and using $F(r_*)=-U^\prime(r_*)$ and that $k$ is an even integer, we obtain equation~(\ref{eqck2}). Thus, we have arrived at the the expressions for $C_{k,2}$ and $C_{k,2}$ that are used in equation~(\ref{taylorck1ck2}).

\section{\label{appb} Thermostats used in MD simulations}

\vskip -3mm

\noindent
Considering 3D simulations in Section~\ref{NumSim}, we use a Nos\'e-Hoover thermostat.
Its parameter, originally~\cite{Nose:1984} denoted Q, is the relaxation time of the thermostat. It is a measure of how strongly the thermostat is attached to the dynamics of the system. We choose a cautious value of $Q=10 \, T$ for each simulation; this linear scaling with $T$ is necessary as we need to more tightly couple the thermostat at lower temperatures in order to accurately maintain the system in the canonical ensemble~\cite{Hunenberger:2005}.

For 1D simulations in Section \ref{1DTheory}, we maintain the canonical ensemble at a target (reduced) temperature $T$ by implementing a Langevin thermostat. This is due to problems with ergodicity utilising the Nosé-Hoover thermostat for small systems~\cite{Tuckerman:2001,Tupper:2005}. Here the evolution of the free particle is modelled (in reduced units) as~\cite{Schlick:2002,Leimkuhler:2015:MDD}
\begin{equation}
\ddot{x}=-\frac{\mbox{d}U}{\mbox{d}x} - \gamma \, \dot{x} +\sqrt{2 \, \gamma \, T} \, R(t) \,,
\label{langevin}
\end{equation}
where $R(t)$ is standard white noise, and $\gamma$ acts as a friction parameter. We choose $\gamma=0.1$ when calculating our illustrative results presented Figures~\ref{1Da4} and~\ref{laplacetemp1d}.

\section*{\label{References} References}

\end{document}